\documentclass[final]{aipproc}

\layoutstyle{6x9}
\newcommand{\msun}{M$_\odot$}

\begin{document}

\title{Nuclear Ashes: Reviewing Thirty Years of Nucleosynthesis in Classical Novae}

\author{Jordi Jos\'e}{
  address={Dept. F\'{\i}sica i Enginyeria Nuclear, Universitat
 Polit\`{e}cnica de Catalunya, and 
Institut d'Estudis Espacials de Catalunya (IEEC/UPC), Barcelona, Spain}}

\begin{abstract}
One of the observational evidences in support of the {\it thermonuclear runaway model}
for the classical nova outburst relies on the accompanying nucleosynthesis. In this paper, we stress the 
relevant role played by nucleosynthesis in our understanding of the nova phenomenon by constraining models 
through a comparison with both 
the atomic abundance determinations from the ejecta and the isotopic 
ratios measured in presolar grains of a likely nova origin. 
Furthermore, the endpoint of nova nucleosynthesis provides hints for the understanding of the mixing process responsible for the enhanced metallicities found in the ejecta, and reveals also information on the properties of the underlying white dwarf (mass, luminosity...). 

We discuss first the interplay between nova outbursts and the Galactic chemical abundances: Classical nova outbursts are expected to be the major source of $^{13}$C, $^{15}$N and $^{17}$O in the Galaxy, and to contribute to the abundances of other species with $\rm A < 40$, such as $^7$Li or $^{26}$Al. We describe the main nuclear path during the course of the explosion, with special emphasis on the synthesis of
radioactive species, of particular interest for the gamma-ray output predicted  from novae ($^7$Li, $^{18}$F, $^{22}$Na, $^{26}$Al).
An overview of the recent discovery of presolar nova candidate grains, as well as a discussion of the role played by nuclear uncertainties associated
with key reactions of the NeNa-MgAl and Si-Ca regions, are also given.
\end{abstract}

\maketitle

\begin{table}
\begin{tabular}{lll}
\hline
  \tablehead{1}{c}{b}{Reference}
  & \tablehead{1}{c}{b}{Model category}
  & \tablehead{1}{c}{b}{Range of nuclei} \\
\hline
Arnould \& N{\o}rgaard (1975) \cite{Arn75}   & Parametric, 1 zone     & $^3$He, $^7$Li, B, C, N, O \\
Arnould et al. (1980) \cite{Arn80}           & Parametric, 1 zone     & H-Ar \\
Boffin et al. (1993) \cite{Bof93}             & Parametric, 1 \& 2 zones & $^3$He, $^7$Be, $^7$Li, $^8$B, $^9$C \\
Coc et al. (1995) \cite{Coc95}                & Semianalytic, 1 zone      & H-K \\
Coc et al. (2000) \cite{Coc00}                & Hydrodynamic, 1D          & O, F\\
Glasner et al. (1997) \cite{Gla97}            & Hydrodynamic, 2D          & H-F\\
Hernanz et al. (1996) \cite{Her96}            & Hydrodynamic, 1D          & $^3$He, $^7$Be, $^7$Li \\
Hernanz et al. (1999) \cite{Her99}            & Hydrodynamic, 1D          & $^{13}$N, $^{18}$F\\
Hillebrandt \& Thielemann (1982) \cite{Hil82} & Parametric, 1 zone        & H-Ar \\
Iliadis et al. (1999) \cite{Ili99}            & Parametric, 1 zone        & H-Ca \\
Jos\'e et al. (1997) \cite{Jos97a}            & Hydrodynamic, 1D          & Na, Mg, Al \\
Jos\'e \& Hernanz (1997) \cite{Jos97b}        & Hydrodynamic, 1D          & H-Ca \\
Jos\'e \& Hernanz (1998) \cite{Jos98}         & Hydrodynamic, 1D          & H-Ca \\
Jos\'e et al. (1999) \cite{Jos99}             & Hydrodynamic, 1D          & Ne-Na, Mg-Al \\
Jos\'e et al. (2001a) \cite{Jos01a}           & Hydrodynamic, 1D          & $^{13}$N, $^{18}$F, $^7$Be, $^{22}$Na, $^{26}$Al \\
Jos\'e et al. (2001b) \cite{Jos01b}           & Hydrodynamic, 1D          & C,N,O,Al,Mg,Si,Ne  \\
Jos\'e et al. (2001c) \cite{Jos01c}           & Hydrodynamic, 1D          & Si-Ca  \\
Kercek et al. (1998) \cite{Ker98}             & Hydrodynamic, 2D          & H-F\\
Kercek et al. (1999) \cite{Ker99}             & Hydrodynamic, 3D          & H-F\\
Kolb \& Politano (1997) \cite{Kol97}          & Hydrodynamic, 1D          & $^{22}$Na, $^{26}$Al  \\
Kovetz \& Prialnik (1985) \cite{Kov85}        & Hydrodynamic, 1D          & H-O\\
Kovetz \& Prialnik (1997) \cite{Kov97}        & Hydrodynamic, 1D          & H-O\\
Kudryashov \& Tutukov (1995) \cite{Kud95}     & Parametric, 1 zone        & H-Ar \\
Kudryashov et al. (2000) \cite{Kud00}         & Parametric, 1 zone        & H-Ar \\
Lazareff et al. (1979) \cite{Laz79}           & Parametric, 2 zones       & C, N, O, F, Ne\\
Nofar et al. (1991) \cite{Nof91}              & Parametric, 1 zone        & H-Al \\
Politano et al. (1995) \cite{Pol95}           & Hydrodynamic, 1D          & H-Ca \\
Prialnik et al. (1978) \cite{Pri78}           & Hydrodynamic, 1D          & H-O\\
Prialnik et al. (1979) \cite{Pri79}           & Hydrodynamic, 1D          & H-O\\
Prialnik (1986) \cite{Pri86a}                 & Hydrodynamic, 1D          & H-O\\
Prialnik \& Shara (1986) \cite{Pri86b}        & Hydrodynamic, 1D          & H-Ne\\
Prialnik \& Shara (1995) \cite{Pri95}         & Hydrodynamic, 1D          & H-P\\
Shara \& Prialnik (1994) \cite{Sha94}         & Hydrodynamic, 1D          & H-Mg\\
Sparks et al. (1978) \cite{Spa78}             & Hydrodynamic, 1D          & H-O\\
Starrfield et al. (1972) \cite{Sta72}         & Hydrodynamic, 1D          & H-O\\
Starrfield et al. (1974a) \cite{Sta74a}       & Hydrodynamic, 1D          & H-O\\
Starrfield et al. (1974b) \cite{Sta74b}       & Hydrodynamic, 1D          & H-O\\
Starrfield et al. (1978a) \cite{Sta78a}       & Hydrodynamic, 1D          & $^3$He, $^7$Li, C, N, O\\
Starrfield et al. (1978b) \cite{Sta78b}       & Hydrodynamic, 1D          & $^7$Li, C, N, O\\
Starrfield et al. (1992) \cite{Sta92}         & Hydrodynamic, 1D          & H-Ar \\
Starrfield et al. (1993) \cite{Sta93}         & Hydrodynamic, 1D          & H-Ar \\
Starrfield et al. (1998) \cite{Sta98}         & Hydrodynamic, 1D          & H-Ar \\
Starrfield et al. (2000) \cite{Sta00}         & Hydrodynamic, 1D          & H-Ar \\
Starrfield et al. (2001) \cite{Sta01}         & Hydrodynamic, 1D          & H-S \\
Vangioni-Flam et al. (1980) \cite{Van80}      & Parametric, 1 zone        & Ne, Al \\
Wallace \& Woosley (1981) \cite{Wal81}        & Parametric, 1 zone        & H-Al \\
Wanajo et al. (1999) \cite{Wan99}             & Semianalytic, 1 zone      & H-Ca \\
Weiss \& Truran (1990) \cite{Wei90}           & Parametric, 1 zone        & H-Ca \\
Wiescher et al. (1986) \cite{Wie86}           & Parametric, 1 zone        & H-Ar \\
\hline
\end{tabular}
\caption{A sample of publications in refereed journals addressing nucleosynthesis in classical novae}
\label{tab:a}
\end{table}

\section{Galactic alchemy: the interplay between nova outbursts and the 
  Galactic abundances}

 The high peak temperatures achieved during nova explosions, T$_{peak} 
 \sim (2-3) \times 10^8$ K, suggest that
 abundance levels of the intermediate-mass elements in the ejecta must be
 significantly enhanced, as confirmed by spectroscopic determinations in
 well-observed nova shells. This raises the issue of the potential contribution of novae
 to the Galactic abundances, which can be roughly estimated as the product of 
 the Galactic nova rate, the average ejected mass per nova outburst, and
 the Galaxy's lifetime. This order of magnitude estimate points out that novae
 scarcely contribute to the Galaxy's overall metallicity (as compared with
 other major sources, such as supernova explosions), nevertheless 
 they can substantially contribute to the synthesis of some
 largely overproduced species (see Table 1, for a sample of publications    
 addressing nucleosynthesis in classical novae). 
 Hence, classical novae are likely sites for the
  synthesis of most of the Galactic $^{13}$C, $^{15}$N and $^{17}$O, whereas   
they can partially contribute to the Galactic abundances of other species
with $\rm A < 40$, such as $^{7}$Li, $^{19}$F, or $^{26}$Al \cite{Sta98,Jos98}.

\begin{figure}
 \includegraphics[height=.4\textheight,clip=]{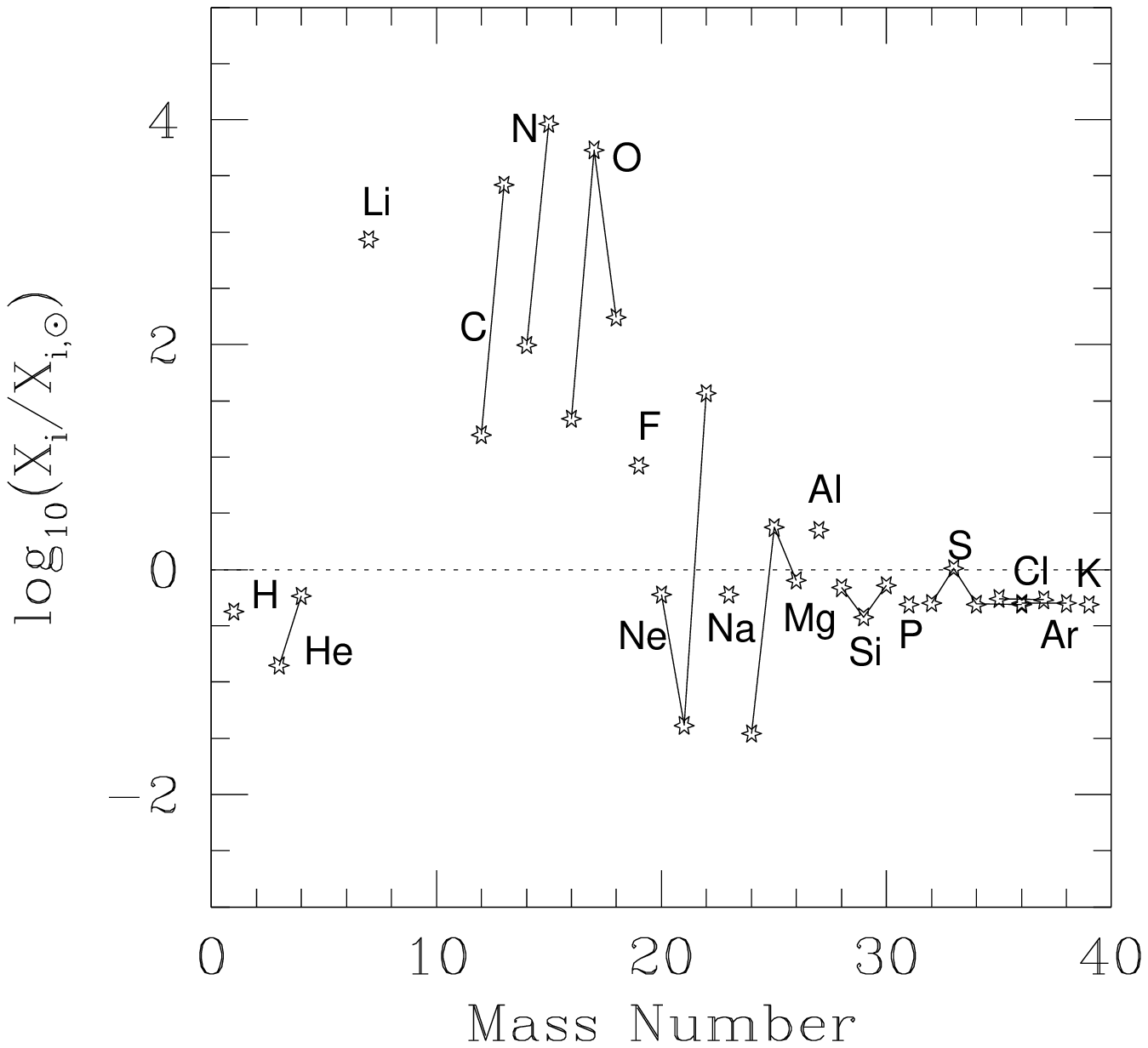}
 \includegraphics[height=.4\textheight,clip=]{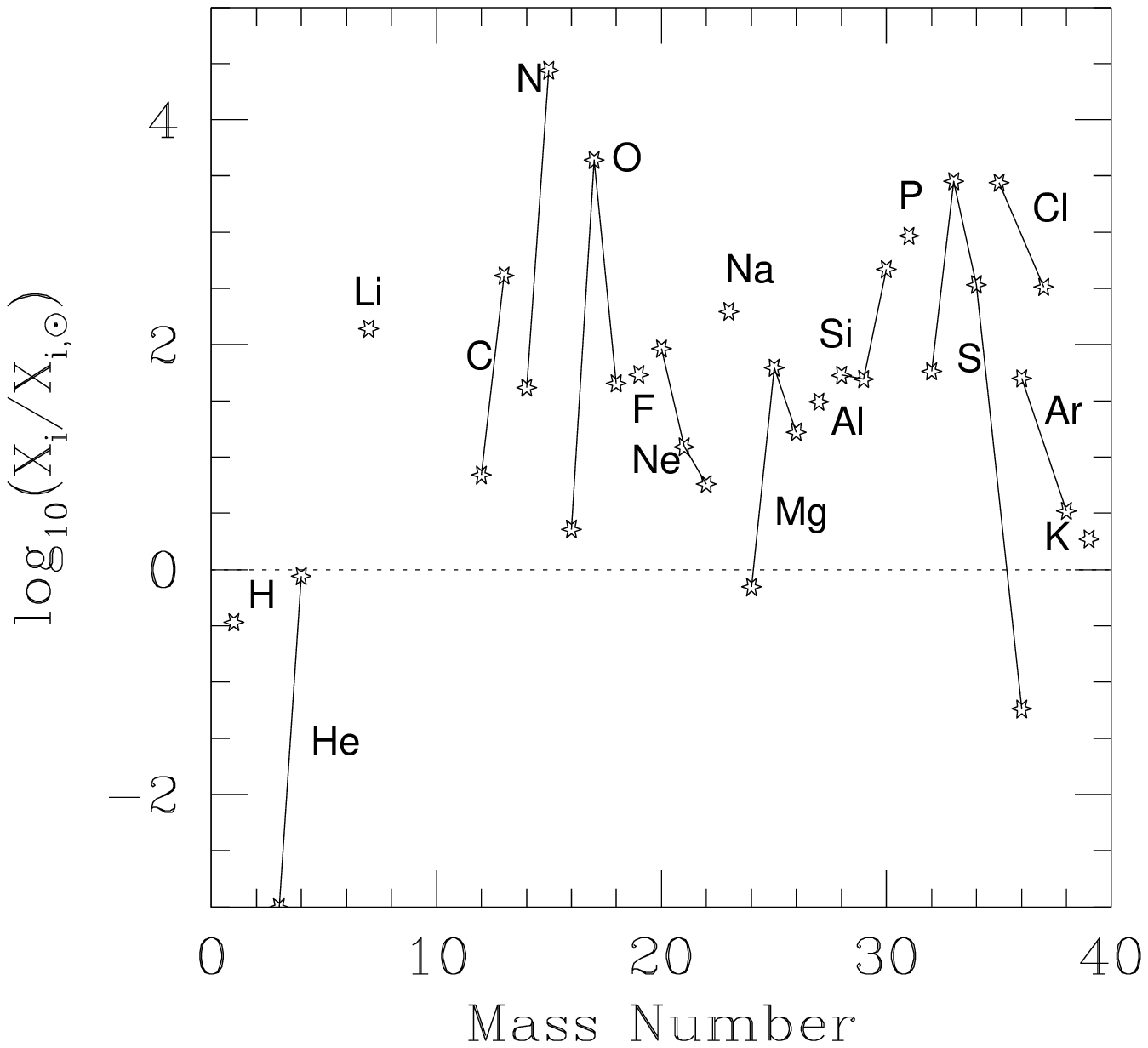}
 \caption{(Left) Mean overproduction factors, relative to solar,  in the
          ejecta of a 1.15 M$_\odot$ CO novae. 
           (Right) Same for a 1.35 M$_\odot$ ONe novae. Figures are based on 
          hydrodynamic calculations reported in \cite{Jos98}.}
\end{figure}

Overproduction factors, relative to solar, corresponding to 
hydrodynamic calculations of nova outbursts on top of a 1.15 M$_\odot$ CO and a 1.35 M$_\odot$ ONe white
 dwarf, are shown in Figure 1. Because of the lower peak temperatures achieved in CO models, and also because of the lack of significant amounts of seed nuclei in the NeNa-MgAl region, the main nuclear activity in CO novae does not extend much beyond oxygen, as seen from the overproduction plot. In contrast, ONe models show a much larger nuclear 
 activity, extending up to silicon (1.15 M$_\odot$ ONe) or argon (1.35 M$_\odot$ ONe).  Hence, the presence of significantly large amounts of intermediate-mass nuclei in the spectra, such as phosphorus,
 sulfur, chlorine or argon, may reveal the presence of an underlying 
 massive ONe white dwarf. Another trend derived from the analysis of the nucleosynthesis
 accompanying nova outbursts is the fact that the O/N and C/N ratios decrease as the mass of the white dwarf (and hence, the peak temperature attained during the 
 explosion) increases.

\subsection{Abundance Determinations in the Ejecta from Novae}

In order to constraint the models, several works have focused on a direct comparison of the atomic 
abundances inferred from observations of the ejecta with the theoretical nucleosynthetic output 
(see \cite{Jos98,Sta98}, and 
references therein). Despite of the problems associated with the modeling of the 
explosion \cite{Sta02}, such as the unknown mechanism responsible for the mixing between the accreted 
envelope and the outermost shells of the underlying white dwarf \cite{Cal02,Dur02}, or the 
difficulties to eject as much material as inferred from observations \cite{Sho02}, there is an excellent 
agreement between theory and observations as regards nucleosynthesis 
(i.e.,  including atomic abundances -H, He, C, O, Ne, Na-Fe-, and a plausible endpoint for nova 
nucleosynthesis). 
In some cases, such as for PW Vul 1984, the agreement between observations and theoretical predictions 
(see \cite{Jos98}, Table 5, for details) is really overwhelming. The reader is referred to \cite{Geh98} for 
an extended list of abundance determinations in the ejecta from novae, and to \cite{Sch02, Van02} for 
recent efforts to improve the abundance pattern for QU Vul 1984 and V1974 Cyg 1992, respectively.

Since the nuclear path is very sensitive to details of the evolution (chemical composition, 
extend of convective mixing, thermal history of the envelope...), the agreement between 
inferred abundances and theoretical yields not only validates the thermonuclear runaway model, 
but also poses 
limits on the (yet unknown) mixing mechanism itself: for instance, if 
mixing occurs very late in the course of the explosion, the accumulation of larger amounts of matter in the envelope will be favored (since the injection of significant amounts of the triggering nucleus $^{12}$C 
will be delayed). Hence, one would expect to end up with a more violent outburst, characterized by a higher T$_{peak}$, exceeding in some cases $4 \times 10^8$ K, and, as a result, a significant enrichment in heavier species, beyond calcium, in the ejecta from novae involving very massive white dwarfs, a pattern never observed so far. 

\subsection{Presolar Grains: Gifts from Heaven}

Infrared \cite{Eva90,Geh98} and ultraviolet 
observations \cite{Sho94} of the temporal evolution of nova light curves
suggest that  novae form grains in the expanding nova shells.
Both CO and ONe novae behave similarly in the infrared right after the
outburst. However, as the ejected envelope expands and becomes optically thin, 
such behavior dramatically changes: CO novae are typically followed by a phase of dust formation corresponding
to a decline in visual light, together with a simultaneous rise in the
infrared emission \cite{Eva02,Geh02}. In contrast, it has been argued
that ONe novae (that involve more massive white dwarfs than CO novae) are not
so prolific producers of dust as a result of the lower mass, high-velocity
ejecta, where the typical densities can be low enough to enable the
condensation of appreciable amounts of dust. Hints on the condensation of dust containing silicates, silicon carbide, carbon and hydrocarbons have been reported from a number of novae (see \cite{Geh98} for a recent review). 

Up to now, the identification of presolar nova grains, presumably condensed in
the shells ejected during the explosion, relied only on low
$^{20}$Ne/$^{22}$Ne ratios (attributed to $^{22}$Na decay), but quite recently
five silicon carbide and two graphite grains that exhibit  
isotopic signatures characteristic of nova nucleosynthesis have been identified \cite{Ama01,Ama02}. 
They are characterized by very low $^{12}$C/$^{13}$C
and $^{14}$N/$^{15}$N ratios,  $^{30}$Si excesses and close-to- or slightly 
lower-than-solar $^{29}$Si/$^{28}$Si ratios, high $^{26}$Al/$^{27}$Al ratios 
(determined only for two grains) and low $^{20}$Ne/$^{22}$Ne ratios 
(only measured in the graphite grain KFB1a-161). Such a promising discovery provides a much valuable source of constraint for nova nucleosynthesis 
(since contrary to the atomic abundance determinations derived from nova ejecta, measurements provide 
more accurate isotopic ratios) and opens interesting possibilities for the future.  

Theoretical isotopic ratios for a variety of nuclear species, ranging from C to Si, based on hydrodynamic computations of the nova outburst, have been
reported by different authors \cite{Sta97,Jos01b,Jos01d,Jos02a}. A more detailed analysis, which
focuses on the different chemical pattern expected for CO and ONe novae, will be presented elsewhere \cite{Jos02b}. 

\section{Synthesis of radioactive nuclei during nova outbursts}

 Among the isotopes synthesized during classical nova outbursts, several
 radioactive species deserve a particular attention. Short-lived nuclei,
 such as $^{14, 15}$O and $^{17}$F (and to some extent $^{13}$N) have 
 been identified as the key isotopes that power the expansion and further
 ejection during a nova outburst through a sudden release of energy, a few
 minutes after peak temperature \cite{Sta72}. 
 Other isotopes have been extensively investigated in connection with the  
 theoretical gamma-ray output from novae \cite{Cla74,Cla81,Lei87}. Hence,
 $^{13}$N and $^{18}$F are responsible for the predicted prompt $\gamma$-ray
 emission \cite{Her99} at and below 511 keV, whereas 
 $^7$Be and $^{22}$Na \cite{Gom98}, which decay much later, when the
 envelope is optically thin, are the sources that power line emission at 478 
 and 1275 keV, respectively. $^{26}$Al is another 
 important radioactive isotope that can be synthesized during nova outbursts, 
 although only its cumulative emission can be observed because of its slow 
 decay.

 We will briefly focus on the corresponding nuclear
 paths leading to the synthesis of the abovementioned gamma-ray emitters,
 with special emphasis on the nuclear uncertainties associated 
 with the relevant reaction rates. We refer the reader to {\cite{Her02a} 
 for a review of the
 current theoretical predictions of the gamma-ray output from 
 novae and of the chances for a nearby future detection using spacecrafts
 such as INTEGRAL (see also \cite{Her02b}). As for a comprehensive 
 summary of the main uncertainties affecting nuclear reaction rates for
 nova temperatures, the reader is referred to \cite{Coc01}. Other recent
 attempts to fully analyze the impact of nuclear physics uncertainties
 in nova nucleosynthesis involve parametric one-zone calculations with
 temperature and density profiles extracted from hydrodynamic models 
 \cite{Ili02a,Ili02b}, as well as Monte Carlo simulations \cite{Smi02}. 
 We would like to stress, however, that results based on parametric
  calculations usually tend to overestimate the effect of a given 
  reaction rate uncertainty, as compared with the outcome from 
 hydrodynamic tests. Hence, although these simplified techniques 
 provide a very valuable tool
 to potentially identify key reactions through an extraordinary large
 number of tests, final conclusions have to be confirmed through
 hydrodynamic calculations.

\subsection{$^7$Be-$^7$Li}

 $^3$He($\alpha,\gamma$)$^7$Be is the main reaction leading to $^7$Be synthesis
(with $^3$He coming from both the accreted amount plus the contribution in 
 place from $^1$H(p,e$^+$ $\nu_e$)$^2$H(p,$\gamma$)$^3$He), which is transformed into $^7$Li by means of an electron capture, emitting a $\gamma$-ray photon of
 478 keV \cite{Gom98}.

 Its production in classical nova outbursts has been very controversial. Results from the first 
pioneering calculation, in the framework of a simple parametric model \cite{Arn75}, were confirmed by 
early hydrodynamic simulations \cite{Sta78a}, assuming however 
an envelope in-place (the first hydrodynamic nova models that properly included 
the onset of the accretion phase were not available until the 80s), thus neglecting the impact of the accretion phase on the evolution. These results were refuted later on, in terms of parametric one/two zone models \cite{Bof93}, pointing out the critical role played by the $^9$C(p,$\gamma$) reaction, not included in 
all previous works (i.e., \cite{Arn75,Sta78a}), and claiming therefore for an unlikely synthesis of $^7$Li in novae.
The scenario was recently revisited by \cite{Her96,Jos98}, who performed new hydrodynamic calculations, 
taking into account both the accretion and explosion stages, and a full reaction network 
(including $^9$C(p,$\gamma$)). These studies confirmed that the Be-transport 
mechanism \cite{Cam55} is able to produce a large overproduction of $^7$Li in nova explosions. 
 
 Among the issues that affect $^7$Li synthesis in novae, one the most critical
 ones is the final amount of $^3$He that survives the early rise in 
 temperature when the thermonuclear runaway ensues. In particular, 
 the different timescales to reach $T_{peak}$ achieved for CO and ONe novae, 
 which deeply depend on the initial $^{12}$C content in the envelope, lead to a 
 larger amount of $^{7}$Be in CO novae (which survives destruction through
 $^7$Be(p,$\gamma$)$^8$B because of the very efficient inverse photodisintegration 
 reaction on $^8$B).  No relevant nuclear uncertainties in the domain
 of nova temperatures affect the corresponding reaction rates. 

The puzzling $^7$Li synthesis in novae, 
suggested for many years but elusive up to now, has been apparently confirmed for the first time 
in the spectra of V382 Vel (Nova Velorum 1999) \cite{Del02}, for which 
 an observed feature compatible with the doublet at 6708 \AA \, of LiI has been reported. Although this 
identification is not yet confirmed (there is no unambiguous interpretation for such a feature),
 the lithium abundances inferred are fully compatible with the theoretical upper
limits given for a fast nova (i.e., \cite{Her96,Jos98}). 

The potential contribution of classical novae to the Galactic 
$^7$Li content turns out to be rather small (i.e., less than 15\%), even if the 
$^7$Li yield from the most favorable case, a massive CO nova \cite{Her96}, is
 considered. Nevertheless, a nova contribution is required to match the $^7$Li 
content in realistic calculations of Galactic chemical evolution 
\cite{Rom99,Rom02}.

\subsection{$^{22}$Na}

The potential role of $^{22}$Na for diagnosis of nova outbursts was first suggested by \cite{Cla74}. It decays to a short-lived excited state of
${}^{22}$Ne (with a lifetime of $\tau = 3.75$ yr), which de-excites to
its ground state by emitting a $\gamma$-ray
photon of $1.275$ MeV. Through this mechanism, nearby ONe novae within a few
kiloparsecs from the Sun may provide detectable $\gamma$-ray fluxes \cite{Her02a}. Several
experimental verifications of this $\gamma$-ray emission at 1.275 MeV from
nearby novae have been attempted in the last twenty years, using balloon-borne
experiments and detectors on-board satellites such as HEAO-3, SMM, or CGRO,
from which upper limits on the ejected $^{22}$Na have been derived.  In 
particular, the observations performed with the COMPTEL
experiment on-board CGRO of five recent Ne-type novae \cite{Iyu95}, as
well as observations of {\it standard} CO novae, have led to an upper limit of
$3.7 \times 10^{-8}$ \msun for the $^{22}$Na mass ejected by any nova in the
Galactic disk. A limit that poses some constraints on pre-existing theoretical models of classical nova explosions.

 Synthesis of $^{22}$Na in novae, extensively investigated in the last two decades \cite{Hil82,Wie86,Woo86,Wei90,Pau91,Pol95,Coc95,Kud95,Kol97,Jos98,Sta98,Jos99,
Wan99,Kud00,Sta00,Jos01a,Sta01}, proceeds through different reaction
 paths. In the $^{20}$Ne-enriched envelopes of ONe novae \cite{Jos99}, it takes place through $^{20}$Ne(p,$\gamma$)$^{21}$Na, followed either by another 
proton capture and a $\beta^+$-decay into $^{22}$Na (i.e., 
 $^{21}$Na(p,$\gamma$)$^{22}$Mg($\beta^+$)$^{22}$Na), or decaying first into 
 $^{21}$Ne before another proton capture ensues (i.e., 
 $^{21}$Na($\beta^+$)$^{21}$Ne(p,$\gamma$)$^{22}$Na). 
Other potential channels, such as proton captures on the seed nucleus $^{23}$Na, play only a marginal role
on $^{22}$Na synthesis because of the much higher initial $^{20}$Ne content in such ONe models.  As for the main destruction channel at nova temperatures, 
 $^{22}$Na(p,$\gamma$)$^{23}$Mg competes favorably with 
 $^{22}$Na($\beta^+$)$^{22}$Ne.
 Nuclear uncertainties strongly affect the rates for 
 $^{21}$Na(p,$\gamma$)$^{22}$Mg and $^{22}$Na(p,$\gamma$)$^{23}$Mg 
 \cite{Jos99}, which translate into an uncertainty
 in the final $^{22}$Na yields (and ultimately on the maximum detectability
 distance of the 1275 keV line expected from nova outbursts). Advances to reduce
the uncertainty affecting the $^{21}$Na(p,$\gamma$) rate have been recently
achieved with the DRAGON recoil separator facility at TRIUMF \cite{Oli02,Bis02}. 

\subsection{$^{26}$Al}

 Several isotopes should be considered as potential seeds for $^{26}$Al
 synthesis: in particular, $^{24,25}$Mg and to some extent $^{23}$Na 
 and $^{20,22}$Ne \cite{Jos97a}. The main nuclear reaction path 
 leading to $^{26}$Al synthesis, also investigated in a large number of papers 
 \cite{Laz79,Arn80,Hil82,Del85,Wie86,Woo86,Wol88,Wei90,Nof91,Pau91,Sha94,
  Pol95,Coc95,Kud95,Kol97,Jos98,Sta98,Jos99,Wan99,Kud00,Sta00,Jos01a,Sta01}, is   
 given by
 $^{24}$Mg(p,$\gamma$)$^{25}$Al($\beta^+$)$^{25}$Mg(p,$\gamma$)$^{26}$Al$^g$,
 whereas destruction is dominated by $^{26}$Al$^g$(p,$\gamma$)$^{26}$Si.

 A significant nuclear uncertainty affects the 
 $^{25}$Al(p,$\gamma$)$^{26}$Si rate \cite{Jos99}, which translates into 
 an uncertainty in the expected contribution of novae to the Galactic 
 $^{26}$Al content. A critical issue in order to estimate this contribution  
is the initial composition of the ONe white dwarf. Whereas calculations by Starrfield et al. 
\cite{Sta86,Sta98,Sta00,Sta01} assume a core composition based on hydrostatic models of carbon burning 
nucleosynthesis \cite{Arn69}, rather enriched in $^{24}$Mg (with a ratio $^{16}$O:$^{20}$Ne:$^{24}$Mg around 
1.5:2.5:1), we adopt more recent values taken from stellar evolution calculations of intermediate-mass stars \cite{Rit96}, for which the $^{24}$Mg content is much smaller ($^{16}$O:$^{20}$Ne:$^{24}$Mg is 10:6:1). Calculations based on the new ONe white dwarf composition \cite{Jos97a,Jos98,Jos99,Jos01a}, 
suggest that the contribution of novae to the Galactic $^{26}$Al abundance 
 is rather small (i.e., less than 15\%), in good agreement with the 
 results derived from the COMPTEL map of the 1809 keV $^{26}$Al emission 
in the Galaxy (see \cite{Die95}), which points towards young progenitors (type 
II supernovae and Wolf-Rayet stars).   

\subsection{$^{18}$F}

 The predicted gamma-ray emission from novae at and below 511 keV at early
 epochs after the explosion is basically driven by the amount of $^{18}$F
 present in the envelope \cite{Lei87,Gom98,Her99}. The synthesis of $^{18}$F is powered by $^{16}$O(p,$\gamma$)$^{17}$F, followed either by 
 $^{17}$F(p,$\gamma$)$^{18}$Ne($\beta^+$)$^{18}$F or by
 $^{17}$F($\beta^+$)$^{17}$O(p,$\gamma$)$^{18}$F. The dominant destruction
 channel is $^{18}$F(p,$\alpha$)$^{15}$O plus a minor contribution from
 $^{18}$F(p,$\gamma$)$^{19}$Ne.
 The effect of the nuclear uncertainties associated with some of the rates
 (i.e., $^{18}$F(p,$\alpha$)$^{15}$O, $^{18}$F(p,$\gamma$)$^{19}$Ne,
  $^{17}$O(p,$\alpha$)$^{14}$N and $^{17}$O(p,$\gamma$)$^{18}$F)
  \cite{Her99,Coc00,Coc01} is, in this case, quite remarkable: they translate into a large uncertainty in the expected $^{18}$F yield, and therefore, in the corresponding gamma-ray flux and maximum detectability distance. 
Advances to reduce
this uncertainty have been reviewed during the Classical Nova Conference 
\cite{Bar02,Des02} and elsewhere \cite{Coc02}, and involve several nuclear physics experiments performed in Oak Ridge (USA) and Orsay (France). 

\subsection{The endpoint of nova  nucleosynthesis}

In agreement with the chemical pattern derived from detailed observations of the ejecta, the theoretical 
endpoint for nova nucleosynthesis is limited to A < 40 (i.e., calcium), in agreement with
current theoretical nucleosynthetic estimates, provided that the 
temperatures attained in the envelope during the explosion remain
limited to T$_{peak} \sim (2-3) \times 10^8$ K. 

The nuclear activity in the Si-Ca region has been scarcely analyzed in detail in 
the context of classical nova outbursts \cite{Sta98,Ili99,Wan99,Jos01c}. It is powered by a leakage from 
the NeNa-MgAl region, where the activity is confined during the early stages of 
the explosion. The main reaction that drives the nuclear activity towards 
heavier species (i.e., beyond S) is mainly $^{30}$P(p,$\gamma$)$^{31}$S, either
followed by $^{31}$S(p,$\gamma$)$^{32}$Cl($\beta^+$)$^{32}$S, or by
$^{31}$S($\beta^+$)$^{31}$P(p,$\gamma$)$^{32}$S \cite{Jos01c}.
The $^{30}$P(p,$\gamma$) rate is based only on Hauser-Feshbach estimates, which can be rather uncertain at 
the domain of nova temperatures. To test the effect of this uncertainty on the predicted yields, we have 
performed a series of hydrodynamic calculations \cite{Jos01c}, modifying arbitrarily the nominal rate. 
Hence, for a high $^{30}$P(p,$\gamma$) rate (i.e., 100 times the nominal one), the final $^{30}$Si yields 
are dramatically reduced by a factor of 30, whereas for a low $^{30}$P(p,$\gamma$) rate (i.e., 0.01 times 
the nominal one), the final $^{30}$Si yields are slightly increased by a factor of 5, whereas isotopes 
above silicon are reduced by a factor of $\sim$ 10,  with dramatic impact on theoretical estimates
ionvolving both the composition of the ejecta and of presolar grains.
\\

The impact of nuclear reaction rate uncertainties on nucleosynthesis calculations points out the need of accurate nuclear physics inputs, and stresses the role played by classical novae as perfect laboratories for nuclear astrophysics.

\begin{theacknowledgments}
First, I would like to thank Margarida Hernanz, for a continuous support. I have learned much from her invaluable criticism, encouragement, and advice. I would also like to thank Sachiko Amari, Alain Coc, John D'Auria, Christian Iliadis, Jordi Isern, Jim MacDonald, Steve Shore, Sumner Starrfield, Jim Truran, Michael Wiescher, and Ernst Zinner,
among others, for many fruitful and stimulating discussions on several aspects involving classical nova outbursts.
\end{theacknowledgments}

\bibliographystyle{aiproc}

\end{document}